\documentclass[12pt]{article}
\usepackage{amsmath,amssymb,bm,graphicx}
\usepackage{color} 	

\newcommand{\delslash}{\not \! \partial}

\begin{document}

\begin{flushright}
\end{flushright}


\begin{center}
{\Large{\bf Parity and CP operations for Majorana neutrinos }}
\end{center}
\vskip .5 truecm
\begin{center}
\bf { Kazuo Fujikawa }
\end{center}

\begin{center}
\vspace*{0.4cm} 
{\it {Interdisciplinary Theoretical and Mathematical Sciences Program,\\
RIKEN, Wako 351-0198, Japan}
}
\end{center}
\makeatletter
\makeatother


\begin{abstract} 
The parity transformation law of  the fermion field $\psi(x)$ is usually defined by  the  "$\gamma^{0}$-parity" $\psi^{p}(t,-\vec{x}) = \gamma^{0}\psi(t,-\vec{x})$ with eigenvalues $\pm 1$, while the "$i\gamma^{0}$-parity" $\psi^{p}(t,-\vec{x})=i\gamma^{0}\psi(t,-\vec{x})$ is required for the Majorana fermion. The compatibility issues of these two parity laws arise in generic fermion number violating theories  where a general class of Majorana fermions appear. 
In the case of  
Majorana neutrinos constructed from  chiral neutrinos in an extension of the Standard Model,  the Majorana neutrinos can be characterized by CP symmetry although C and P are separately broken.  It is then shown that either choice of the parity operation, $\gamma^{0}$ or $i\gamma^{0}$, in the level of the starting fermions  
gives rise to the consistent and physically equivalent descriptions of emergent Majorana neutrinos both for Weinberg's model of neutrinos and for a general class of seesaw models. The mechanism of this equivalence is that the Majorana neutrino constructed from a chiral neutrino, which satisfies the classical Majorana condition $\psi(x)=C\overline{\psi(x)}^{T}$, allows the phase freedom   $\psi(x)=e^{i\alpha}\nu_{L}(x) + e^{-i\alpha}C\overline{\nu_{L}(x)}^{T}$ with $\alpha=0\ {\rm or}\ \pi/4$ that accounts for the phase coming from the different definitions of parity for $\nu_{L}(x)$ and ensures the consistent  definitions of CP symmetry $({\cal CP})\psi(x)({\cal CP})^{\dagger}= \pm i\gamma^{0}\psi(t,-\vec{x})$.
\end{abstract}


\section{Introduction}
The parity symmetry, which is a disconnected component of the Lorentz transformation, is defined for the Dirac fermion with real $m$
\begin{eqnarray}\label{Dirac action}
S=\int d^{4}x \overline{\psi(x)}[i\gamma^{\mu}\partial_{\mu}-m]\psi(x)
\end{eqnarray}
by the substitution rule \cite{Bjorken}
\begin{eqnarray}\label{parity of Dirac}
\psi(x) \rightarrow \gamma^{0}\psi(t, - \vec{x}) = \psi^{p}(t,-\vec{x}), \ \  \psi^{p}(t,-\vec{x}) \rightarrow \psi(x)
\end{eqnarray}
which is the symmetry of \eqref{Dirac action}.
The above parity symmetry P is understood  physically as the mirror symmetry, and thus good parity implies left-right symmetry.
 The charge conjugation symmetry of the Dirac fermion is defined by the representation theory of the Clifford algebra using $C=i\gamma^{2}\gamma^{0}$ in the convention of \cite{Bjorken} by 
\begin{eqnarray}\label{C of Dirac}
 \psi(x) \rightarrow C\overline{\psi(x)}^{T} = \psi^{c}(x), \ \  \psi^{c}(x) \rightarrow C\overline{\psi^{c}(x)} = \psi(x) 
\end{eqnarray}
which is the symmetry of the action \eqref{Dirac action}. 
The combination of C and P then gives \begin{eqnarray}\label{CP1}
&&\psi(x) \rightarrow \psi^{cp}(t, -\vec{x})\equiv (\psi^{c})^{p}(t,-\vec{x}) =C\overline{\gamma^{0}\psi(t,-\vec{x})}
= -\gamma^{0}\psi^{c}(t,-\vec{x}),\nonumber\\
&&\psi(x) \rightarrow \psi^{pc}(t, -\vec{x})\equiv (\psi^{p})^{c}(t,-\vec{x}) =\gamma^{0}C\overline{\psi(t,-\vec{x})}
= \gamma^{0}\psi^{c}(t,-\vec{x})
\end{eqnarray}    
which is also the symmetry of the action \eqref{Dirac action}. As for the ordering of the operation,  we have 
${\cal C} \psi(x){\cal C}^{\dagger} = C\overline{\psi(x)}^{T}$ and ${\cal P} \psi(x){\cal P}^{\dagger} = \gamma^{0}\psi(t,-\vec{x})$ in a formal operator notation, and  thus  
\begin{eqnarray}
&&{\cal P}{\cal C} \psi(x){\cal C}^{\dagger}{\cal P}^{\dagger} = {\cal P}C\overline{\psi(x)}^{T}{\cal P}^{\dagger} = C\overline{\gamma^{0}\psi(t,-\vec{x})}^{T} =
-\gamma^{0}\psi^{c}(t,-\vec{x}) = \psi^{cp}(t, -\vec{x}),\nonumber\\
&&{\cal C}{\cal P} \psi(x){\cal P}^{\dagger}{\cal C}^{\dagger} = {\cal C}\gamma^{0}\psi(t,-\vec{x}){\cal C}^{\dagger} = \gamma^{0}C\overline{\psi(t,-\vec{x})}^{T} =
\gamma^{0}\psi^{c}(t,-\vec{x}) = \psi^{pc}(t, -\vec{x}).
\end{eqnarray} 
This rule shows that the order of C and P is important when one considers CP, although the action $S=\int d^{4}{\cal L}$ which is bilinear in the fermion field is invariant for either way of the combination of C and P if it is invariant for one of the orders. As is well-known, the intrinsic parity of the Dirac fermion $\psi^{p}(t,-\vec{x}) = \gamma^{0}\psi(t,-\vec{x})$ and the intrinsic parity of its antifermion ${(\psi^{c})}^{p}(t,-\vec{x}) = -\gamma^{0}\psi^{c}(t,-\vec{x})$ have opposite signatures and it is physically important to define the parity of a composite boson such as the charmonium.      

On the other hand, the original Majorana fermion is defined by the same action as \eqref{Dirac action}
but with {\em purely imaginary} Dirac gamma matrices $\gamma^{\mu}$ \cite{Majorana}. Then the Dirac equation
\begin{eqnarray}
[i\gamma^{\mu}\partial_{\mu}-m]\psi(x) = 0
\end{eqnarray}
is a real differential equation, and one can impose the reality condition on the solution
\begin{eqnarray}\label{reality condition}
\psi(x)^{\star} = \psi(x)
\end{eqnarray} 
which implies the self-conjugate under the charge conjugation~\footnote{The pure imaginary condition $\psi^{\star}(x)=-\psi(x)$ is also  allowed, but we take \eqref{reality condition} as the primary definition in the present paper. 
}. The conventional parity transformation
$\psi(x) \rightarrow \psi^{p}(t, -\vec{x}) = \gamma^{0}\psi(t, -\vec{x})$ cannot maintain the reality condition \eqref{reality condition} for the purely imaginary $\gamma^{0}$.  Thus the ``$i\gamma^{0}$-parity'' 
\begin{eqnarray}\label{modified parity}
\psi(x) \rightarrow  \psi^{p}(t,-\vec{x}) = i\gamma^{0}\psi(t,-\vec{x})
\end{eqnarray}
is chosen as a natural parity transformation rule for the Majorana fermion \cite{Majorana, Kayser1}.  

In the generic representation of the Dirac matrices \cite{Bjorken}, the ``$i\gamma^{0}$-parity''  satisfies the condition 
\begin{eqnarray}\label{consistency condition}
i\gamma^{0}\psi(t,-\vec{x}) = C\overline{i\gamma^{0}\psi(t,-\vec{x})}^{T}
\end{eqnarray}
for the field which satisfies the classical Majorana condition 
\begin{eqnarray}
\psi (x) = C\overline{\psi(x)}^{T}
\end{eqnarray}
and thus  $i\gamma^{0}$-parity is a natural choice of the parity for the Majorana fermion  in this generic representation  also. The choice $\psi^{p}(t,-\vec{x}) = -i\gamma^{0}\psi(t,-\vec{x})$ also satisfies the condition \eqref{consistency condition}, but we choose the convention $\psi^{p}(t,-\vec{x}) =i\gamma^{0}\psi(t,-\vec{x})$ in \eqref{modified parity}  as a primary definition.  For consistency, we assign the $i\gamma^{0}$-parity convention to charged leptons also when we assign $i\gamma^{0}$-parity to neutrinos, although this extra phase is cancelled in the lepton number conserving terms in the charged lepton sector. See \cite{Weinberg2} for the phase freedom of parity operation.

The combination of C and P for the Majorana fermion then becomes 
\begin{eqnarray}\label{CP2}
&&\psi(x) \rightarrow {\psi^{c}}^{p}(t, -\vec{x})= C\overline{\psi^{p}(t,-\vec{x})}=C\overline{i\gamma^{0}\psi(t,-\vec{x})}
= i\gamma^{0}\psi^{c}(t,-\vec{x}),\nonumber\\
&&\psi(x) \rightarrow {\psi^{p}}^{c}(t, -\vec{x}) = i\gamma^{0}C\overline{\psi(t,-\vec{x})}
= i\gamma^{0}\psi^{c}(t,-\vec{x})
\end{eqnarray}    
which is also the symmetry of the action \eqref{Dirac action}. In the present case, CP operation does not depend on the ordering of C and P. Also, the parity $\psi^{p}(t,-\vec{x}) = i\gamma^{0}\psi(t,-\vec{x})$ of a fermion and the parity of an antifermion ${\psi^{c}}^{p}(t,-\vec{x}) = i\gamma^{0}\psi^{c}(t,-\vec{x})$ have  symmetric
forms, which is natural since we do not distinguish the particle and its antiparticle in the case of the Majorana fermion.

These two choices of parity, $\gamma^{0}$ and $i\gamma^{0}$, are equivalent for the Dirac fermion with the fermion number $U(1)$ freedom since $i\gamma^{0}$ is regarded as a composition of $\gamma^{0}$ and $U(1)$ transformations, but their physical equivalence in generic theories with   fermion number non-conservation  is not obvious. One may rather suspect that the common $\gamma^{0}$-parity is inconsistent in theories where the Majorana fermion appears. 
In the following sections, we are going to show that  these two different choices of the parity, $\gamma^{0}$ and $i\gamma^{0}$, give the physically equivalent descriptions of Majorana neutrinos using Weinberg's model~\cite{Weinberg1} in an extension of the Standard Model. Namely, we show that the different definitions of parity in the level of  starting fermions in Weinberg's model lead to the consistent and physically equivalent descriptions of emergent Majorana neutrinos using the CP symmetry to characterize Majorana neutrinos. The same statement applies to a general class of seesaw models of Majorana neutrinos, which is shown in Appendix. It appears that the conventional  $\gamma^{0}$-parity is often  used in the phenomenological analyses of the models with Majorana neutrinos~\cite{Xing}, while theoretically the use of $i\gamma^{0}$-parity is  natural.  An explicit demonstration of the physical equivalence of the two different definitions of parity in the analysis of Majorana neutrinos in an extension of SM and the explanation of its basic mechanism will thus be practically useful.

Technically,  we employ the characterization of Majorana fermions constructed from chiral fermions by the CP symmetry as a basic means to discuss this issue.
The characterization of the Majorana fermion such as $\psi(x) = \nu_{L}(x) + C\overline{\nu_{L}}^{T}(x)$ by the CP symmetry has been suggested in \cite{Fujikawa} since this field has no symmetry under P nor C but can have a good symmetry under the CP symmetry, as will be discussed later. In this sense, our characterization  of the Majorana neutrino in an extension of the Standard Model  by the CP symmetry is very close to the characterization of the Weyl neutrino by the CP symmetry in SM.  It should be noted that 
our use of the CP symmetry for the Majorana fermion, which is constructed from chiral fermions, is very different from the proposal of the use of CP or ultimately CPT symmetries in the definition of  general Majorana neutrinos by taking into account of  the possible symmetry breaking by weak interactions, as was discussed, for example, in \cite{Kayser2}.

\section{Weinberg's model of massive Majorana neutrinos}

Weinberg's model of massive Majorana neutrinos in an extension of the Standard Model  is defined in terms of chiral fermions~\cite{Weinberg1}.  It is known that Weinberg's model of Majorana neutrinos represent the essential aspects of the general models of Majorana neutrinos such as the seesaw models \cite{Xing, Fukugita, Giunti, Bilenky}. We are going to compare the emergent Majorana neutrinos when we use the two different definitions of parity operation for starting fermions in Weinberg's model.

\subsection{The $i\gamma^{0}$-parity}
We define the charge conjugation and parity operations of chiral fermions by the chiral projection of the transformation rules of the Dirac fermion using the operator notation 
\begin{eqnarray}\label{conventional C and P}
&&{\cal C}\nu_{L}(x){\cal C}^{\dagger}=C\overline{\nu_{R}(x)}^{T},\ \ \ {\cal C}\nu_{R}(x){\cal C}^{\dagger}=C\overline{\nu_{L}(x)}^{T},\nonumber\\
&&{\cal P}\nu_{L}(x){\cal P}^{\dagger}=i\gamma^{0}\nu_{R}(t,-\vec{x}),\ \ \ {\cal P}\nu_{R}(x){\cal P}^{\dagger}=i\gamma^{0}\nu_{L}(t,-\vec{x}),\nonumber\\
&&({\cal P}{\cal C})\nu_{L}(x)({\cal P}{\cal C})^{\dagger}=i\gamma^{0}C\overline{\nu_{L}(t,-\vec{x})}^{T},\ \ \ ({\cal P}{\cal C})\nu_{R}(x)({\cal P}{\cal C})^{\dagger}=i\gamma^{0}C\overline{\nu_{R}(t,-\vec{x})}^{T}
\end{eqnarray}  
where we adopt the $i\gamma^{0}$-parity ${\cal P}\nu(x){\cal P}^{\dagger}=i\gamma^{0}\nu(t,-\vec{x})$ for a Dirac fermion with
\begin{eqnarray}\label{chiral projection}
\nu_{R,L}(x)=(\frac{1\pm \gamma_{5}}{2})\nu(x).
\end{eqnarray}
The relations \eqref{conventional C and P}  fix the notational conventions of the transformation laws of chiral fermions. We are not assuming that the actual massive neutrinos are Dirac fermions, although we use the notations $\nu_{L,R}(x)$ for simplicity. 
These rules extracted from the Dirac fermion are mathematically consistent and the symmetries of the action 
\begin{eqnarray}\label{left-right symmetric action} 
S &=& \int d^{4}\{\overline{\nu_{L}}(x)i\gamma^{\mu}\partial_{\mu}\nu_{L}(x) + \overline{\nu_{R}}(x)i\gamma^{\mu}\partial_{\mu}\nu_{R}(x)\nonumber\\
&& - m\overline{\nu_{L}}(x)\nu_{R}(x)
- m\overline{\nu_{R}}(x)\nu_{L}(x)\}.
\end{eqnarray}
Physically, parity is defined as the mirror symmetry and one can check if the given Lagrangian is parity preserving or not using these rules. 
Good P naturally implies left-right symmetry, and P is represented in the form of a doublet representation $\{\nu_{R}(x),\nu_{L}(x)\}$. The doublet representation of the charge conjugation C in \eqref{conventional C and P} is related to the absence of the Majorana-Weyl fermion in $d=4$ dimensions, which is a consequence of the representation theory of the Clifford algebra; intuitively, the absence of the Majorana-Weyl fermion is related to the fact that the charge conjugation inevitably changes the signature of 
$\gamma_{5}$ in $d=4$, namely, $\gamma_{5} \rightarrow -\gamma_{5}$ and thus $\nu_{L,R}(x)\rightarrow C\overline{\nu_{R,L}(x)}^{T}$ as in the first line of \eqref{conventional C and P}. If one uses the charge conjugation operation other than \eqref{conventional C and P}, one would have a potential danger of spoiling the condition of the absence of Majorana-Weyl fermions in $d=4$ \cite{Fujikawa}. The definitions of C and P transformation rules of chiral fermions \eqref{conventional C and P} are highly unique in this sense. See \cite{FT1} for the problematic aspects of the ``pseudo C-symmetry'' for the chiral fermion, which is often used in the analysis of seesaw models.  See also \eqref{pseudo operators} later.

Weinberg's model of Majorana neutrinos is defined by an effective  hermitian Lagrangian~\cite{Weinberg1} 
\begin{eqnarray}\label{Weinberg-type}
{\cal L}&=&\overline{\nu_{L}}(x)i\delslash \nu_{L}(x)
-(1/2)\{\nu_{L}^{T}(x)Cm_{L}\nu_{L}(x) + h.c.\}
\end{eqnarray}
with an arbitrary  $3\times 3$ symmetric complex mass matrix $m_{L}$;  $m_{L}$ is symmetric because of the symmetry properties of the matrix $C$ and fermion fields $\nu_{L}(x)$. This Lagrangian contains only the left-handed chiral components and is not invariant under C nor P in \eqref{conventional C and P}. Under the CP transformation in \eqref{conventional C and P}, the action defined by ${\cal L}$ \eqref{Weinberg-type} is transformed as
\begin{eqnarray}
\int d^{4}x{\cal L}&\rightarrow &
\int d^{4}x \{\overline{\nu_{L}}(x)i\delslash \nu_{L}(x)
-(1/2)[\nu_{L}^{T}(x)Cm^{\dagger}_{L}\nu_{L}(x) + h.c.]\}
\end{eqnarray}
and thus ${\cal L}$ is CP invariant (in terms of $i\gamma^{0}$ parity) if the symmetric mass matrix is real 
\begin{eqnarray}\label{real mass condition}
m^{\dagger}_{L} = m_{L}.
\end{eqnarray}
To be precise, we use this CP invariance condition 
for the diagonalized mass eigenvalues, while leaving the bare parameter $m_{L}$ to be general and not necessarily real.    
After the diagonalization of the symmetric complex mass matrix by the $3\times 3$ Autonne-Takagi factorization using a unitary $U$ \cite{Autonne,Takagi}
\begin{eqnarray}\label{characteristic value1}
U^{T}m_{L}U= M
\end{eqnarray}
with a real $3\times 3$ diagonal matrix $M$, we define 
\begin{eqnarray}\label{A-T transformation1}
\nu_{L}(x)=U\tilde{\nu}_{L}(x)
\end{eqnarray}
and thus transfer the possible CP breaking contained in $U$ to the PMNS mixing matrix which contains a mixing matrix coming from the charged lepton sector also in an extension of the Standard Model.
We then have a hermitian Lagrangian (suppressing the tilde-symbol of $\tilde{\nu}_{L}(x)$)
\begin{eqnarray}\label{Weinberg-type2}
{\cal L}
&=&\overline{\nu_{L}}(x)i\delslash \nu_{L}(x)
-(1/2)\{\nu_{L}^{T}(x)CM\nu_{L}(x) + h.c.\}\nonumber\\
&=&(1/2)\{\overline{\psi}(x)i\delslash \psi(x)- \overline{\psi}(x)M\psi(x)\}          
\end{eqnarray}
where we defined
\begin{eqnarray}\label{Majorana in Weinberg1}
\psi(x)\equiv \nu_{L}(x)+ C\overline{\nu_{L}}^{T}(x).
\end{eqnarray}
The field $\psi(x)$ satisfies the classical Majorana condition 
\begin{eqnarray}\label{classical condition1}
\psi(x)=C\overline{\psi(x)}^{T}
\end{eqnarray}
identically regardless of the choice of $\nu_{L}(x)$  in the sense that the condition \eqref{classical condition1} is satisfied even if one replaces $\nu_{L}(x)$ by an arbitrary chiral $N_{L}(x)$.
In the present approach where the charge conjugation C is not necessarily a good symmetry, we define the Majorana fermion by \eqref{classical condition1} together with the Dirac equation $[i\delslash- M]\psi(x)=0$. The condition \eqref{classical condition1}, which is fixed by the analysis of the Clifford algebra, is referred to as the classical Majorana condition in the present paper.

The transformation \eqref{A-T transformation1} belongs to a {\em canonical transformation} which preserves the form of the kinetic term in the Lagrangian and thus preserves the canonical anti-commutation relations~\cite{Pauli, Gursey, KF-PG}. The well-known Kobayashi-Maskawa analysis is also an example of the use of the canonical transformation \cite{KM}. In the canonical transformation, we apply the transformation rules of discrete symmetries  in \eqref{conventional C and P} to the new variables every time after the canonical transformation. We apply the discrete symmetries in \eqref{conventional C and P} to the old variables also. Thus the discrete symmetries applied to the old variables do not generate the discrete symmetries of the new variables in general~\cite{KF-PG}. This procedure is the same as in the Kobayashi-Maskawa analysis. The Lagrangian in \eqref{Weinberg-type2}, which contains only the left-handed chiral fermions, is thus not invariant under C nor P but invariant under the CP transformation since $M$ is real and diagonal. See the CP invariance condition \eqref{real mass condition} when we define CP in terms of $i\gamma^{0}$ parity.

In the present case \eqref{Weinberg-type2},  C and P are thus not specified for the field $\nu_{L}$, but CP symmetry,
$({\cal P}{\cal C})\nu_{L}(x)({\cal P}{\cal C})^{\dagger}=i\gamma^{0}C\overline{\nu_{L}(t,-\vec{x})}^{T}$, in \eqref{conventional C and P}
is well-defined. The chiral fermion $\nu_{L}(x)$ appearing in $\psi(x)$, of which mass term (the dimension 5 operator) is generated by a renormalization group flow  starting with the massless Weyl fermion in an extension of the Standard Model \cite{Weinberg1}, for example, has well-defined CP after the mass diagonalization  just as the starting massless Weyl fermion in SM.

We thus naturally characterize the Majorana fermion \eqref{Majorana in Weinberg1} by the CP symmetry
\begin{eqnarray}
&&({\cal P}{\cal C})[\nu_{L}(x)+ C\overline{\nu_{L}}^{T}(x)]({\cal P}{\cal C})^{\dagger}=i\gamma^{0}[ C\overline{\nu_{L}}^{T}(t,-\vec{x}) + \nu_{L}(t,-\vec{x})],
\end{eqnarray} 
namely,
\begin{eqnarray}\label{CP to define Majorana}
 ({\cal P}{\cal C})\psi(x)({\cal P}{\cal C})^{\dagger}= i\gamma^{0}C\overline{\psi(t,-\vec{x})}^{T}= i\gamma^{0}\psi(t,-\vec{x}).
\end{eqnarray}
The first equality in \eqref{CP to define Majorana} implies the operator relation while the second equality in \eqref{CP to define Majorana}  implies the classical Majorana condition \eqref{classical condition1} which holds identically in the sense that \eqref{classical condition1} holds irrespective of the choice of $\nu_{L}(x)$.  Note that the CP relation $ \psi(x) \rightarrow i\gamma^{0}\psi(t,-\vec{x})$ in  \eqref{CP to define Majorana}, which is an exact symmetry of \eqref{Weinberg-type2}, preserves the classical Majorana identity  \eqref{classical condition1}
\begin{eqnarray}\label{CP of classical identity1}
 i\gamma^{0}\psi(t,-\vec{x}) = C\overline{ i\gamma^{0}\psi(t,-\vec{x})}^{T},
\end{eqnarray} 
namely, the CP transformation and the classical Majorana condition are consistent. This is an analogue of the crucial consistency of parity operation in \eqref{reality condition} and \eqref{modified parity}.

The chiral component $\nu_{L}(x)$ of $\psi(x)$ describes  the weak interaction in an extension of the Standard Model 
\begin{eqnarray}\label{Weak-coupling1}
\int d^{4}x[ (g/\sqrt{2})\bar{l}_{L}(x)\gamma^{\mu}W_{\mu}(x)U_{PMNS}\nu_{L}(x) + h.c.]
\end{eqnarray}
perfectly well, since the conventional C and P are broken in the parity-violating weak interaction and thus the specification of C and P for $\nu_{L}(x)$ in \eqref{Weak-coupling1} is not required, analogously to the Weyl neutrino in SM. In the present formulation, we use the transformation rules of C, P and CP in \eqref{conventional C and P} for the chiral components of both charged leptons and neutrinos in a uniform manner, although some of them are broken.  We recall that $i\gamma^{0}$-parity rules are applied to charged leptons also when we define $i\gamma^{0}$-parity for neutrinos.

The CP symmetry breaking is described by the PMNS matrix $U_{PMNS}$ when combined with the CP symmetry properties of the charged lepton $l_{L}(x)$ and the chiral component $\nu_{L}(x)$ of the neutrino in \eqref{Weak-coupling1}. The absence of the $U(1)$ phase freedom of $\nu_{L}(x)$ in \eqref{Weinberg-type2}, namely, the lepton number non-conservation,  is important to count an increase in the number of possible CP violating phases in  $U_{PMNS}$; for example, a model with two generations of leptons can have CP violation  \cite{Takasugi}. 
The entire weak interaction is thus described by the chiral component  $\nu_{L}(x)$ using its CP property.
 The Majorana neutrinos characterized by the CP symmetry retain certain information of their original chiral contents; for example,  the parity by itself is not specified in the present characterization of  Majorana neutrinos (the parity is not specified but the CP-parity is specified).
 The fact that the neutrino is a chiral projection of a Majorana fermion  is assured by $\nu_{L}(x)=[(1-\gamma_{5})/2]\psi(x)$ and \eqref{CP to define Majorana} which contains the classical identity \eqref{classical condition1}.
 
 If one wishes to define the parity and charge conjugation operators valid for the Majorana neutrinos in the present formulation,  it is possible to define a {\em deformed symmetry}  generated by  \cite{KF-PG, Fujikawa}
\begin{eqnarray}\label{Deformed C and P} 
  {\cal C}_{M}=1, \ \ \ {\cal P}_{M}={\cal P}{\cal C}
\end{eqnarray}
which is the good symmetry of \eqref{Weinberg-type2}, and the Majorana field $\psi(x)$ is transformed as 
 \begin{eqnarray}\label{trivial C and CP}
{\cal C}_{M}\psi(x){\cal C}_{M}^{\dagger}=\psi(x), \ \ \ {\cal P}_{M}\psi(x){\cal P}_{M}^{\dagger}= i\gamma^{0}\psi(t,-\vec{x}).
\end{eqnarray}
 The non-trivial part of 
 this deformation is the CP symmetry and, in this sense, this deformation is essentially equivalent to the formulation of the Majorana neutrinos with the CP symmetry described above. The classical Majorana condition $\psi(x)=C\overline{\psi(x)}^{T}$, that determines whether  a given fermion is a Majorana fermion or not, carries the same physical information as the trivial operation ${\cal C}_{M}\psi(x){\cal C}_{M}^{\dagger}=\psi(x)$ applied to the fermion $\psi(x)$ which is assumed to be a Majorana fermion $\psi(x)=C\overline{\psi(x)}^{T}$.

The formulation \eqref{trivial C and CP} leads to a formal enhancement of discrete symmetries in  \eqref{Weinberg-type2} by assigning C and P to the chiral component $\nu_{L}(x)=(\frac{1-\gamma_{5}}{2})\psi(x)$,
\begin{eqnarray} \label{Majorana basis1} 
&&{\cal C}_{M}\nu_{L}(x){\cal C}_{M}^{\dagger}=\nu_{L}(x) =C\overline{\nu_{R}}^{T}(x), \nonumber\\ 
&&{\cal P}_{M}\nu_{L}(x){\cal P}_{M}^{\dagger}=i\gamma^{0}C\overline{\nu_{L}(t,-\vec{x})}^{T}=i\gamma^{0}\nu_{R}(t,-\vec{x})
\end{eqnarray}
where we defined $\nu_{R}(x)
\equiv (\frac{1+\gamma_{5}}{2})\psi(x) = C\overline{\nu_{L}}^{T}(x)$, and 
\begin{eqnarray}\label{Majorana basis2}
{\cal C}_{M}\nu_{R}(x){\cal C}_{M}^{\dagger}=C\overline{\nu_{L}}^{T}(x), \ \  {\cal P}_{M}\nu_{R}(x){\cal P}_{M}^{\dagger}=i\gamma^{0}\nu_{L}(t,-\vec{x}).
\end{eqnarray}
These transformation rules are mathematically consistent and imply perfect left-right symmetry  expected for a Majorana fermion $\psi=\nu_{L}+\nu_{R}$.

These rules \eqref{Majorana basis1} and \eqref{Majorana basis2} may be compared to the rules in \eqref{conventional C and P}. The physical degrees of freedom of a Majorana fermion $\psi=\nu_{L}+\nu_{R}$ are the same as either a left-handed chiral fermion or a right-handed chiral fermion but not both, contrary to the case of a Dirac fermion in \eqref{conventional C and P} where the left- and right-handed components are independent.  If one measures the left-handed projection of the Majorana fermion $\psi$, for example, one obtains the chiral freedom $\nu_{L}$ and simultaneously the information of $\nu_{R}$ also.  Physically, this introduction of $ {\cal C}_{M}$ and $ {\cal P}_{M}$ does not add new ingredients to the analysis of weak interactions \cite{Fujikawa}.

\subsection{The $\gamma^{0}$-parity}
We now examine  the same Weinberg's model using the $\gamma^{0}$-parity operation of chiral fermions defined by the chiral projection of the transformation rules of the Dirac fermion, which are written in the operator notation
\begin{eqnarray}\label{conventional C and P-2}
&&{\cal C}\nu_{L}(x){\cal C}^{\dagger}=C\overline{\nu_{R}(x)}^{T},\ \ \ {\cal C}\nu_{R}(x){\cal C}^{\dagger}=C\overline{\nu_{L}(x)}^{T},\nonumber\\
&&{\cal P}\nu_{L}(x){\cal P}^{\dagger}= \gamma^{0}\nu_{R}(t,-\vec{x}),\ \ \ {\cal P}\nu_{R}(x){\cal P}^{\dagger}= \gamma^{0}\nu_{L}(t,-\vec{x}),\nonumber\\
&&({\cal P}{\cal C})\nu_{L}(x)({\cal P}{\cal C})^{\dagger}=-\gamma^{0}C\overline{\nu_{L}(t,-\vec{x})}^{T},\ \ \ ({\cal P}{\cal C})\nu_{R}(x)({\cal P}{\cal C})^{\dagger}=-\gamma^{0}C\overline{\nu_{R}(t,-\vec{x})}^{T}
\end{eqnarray}  
where
\begin{eqnarray}\label{chiral projection}
\nu_{R,L}(x)=(\frac{1\pm \gamma_{5}}{2})\nu(x)
\end{eqnarray}
instead of \eqref{conventional C and P}. Note the appearance of the minus sign in the last two relations in \eqref{conventional C and P-2}, namely, we adopt the convention of $\psi^{cp}(t,-\vec{x})$ in \eqref{CP1}.
These rules extracted from the Dirac fermion are mathematically consistent and the symmetries of the action \eqref{left-right symmetric action}.

Physically, parity is defined as the mirror symmetry. 
Good P naturally implies left-right symmetry, and P is represented in the form of a doublet representation $\{\nu_{R}(x),\nu_{L}(x)\}$. The doublet representation of the charge conjugation C in \eqref{conventional C and P-2} is the same as the charge conjugation C in \eqref{conventional C and P} and it is related to the absence of the Majorana-Weyl fermion in $d=4$ \cite{Fujikawa}.
See \cite{FT1} for the problematic aspects of the ``pseudo C-symmetry'' for the chiral fermion, which is often used in the analysis of seesaw models.  See also \eqref{pseudo operators} later.

Weinberg's model of Majorana neutrinos is defined by the same effective  hermitian Lagrangian as \eqref{Weinberg-type2}
\begin{eqnarray}\label{Weinberg-type3}
{\cal L}&=&\overline{\nu_{L}}(x)i\delslash \nu_{L}(x)
-(1/2)\{ \nu_{L}^{T}(x)Cm_{L}\nu_{L}(x) + \overline{\nu_{L}(x)}Cm^{\dagger}_{L}\overline{\nu_{L}(x)}^{T}\}
\end{eqnarray}
with an arbitrary  $3\times 3$ symmetric complex mass matrix $m_{L}$.
We want to show that the {\em same form} of the starting Lagrangian when analyzed with the  $\gamma^{0}$-parity leads to the logically consistent and  physically equivalent predictions as with the $i\gamma^{0}$-parity.
 
    This Lagrangian \eqref{Weinberg-type3} contains only the left-handed chiral components and is not invariant under C nor P in \eqref{conventional C and P-2}. Under the CP transformation in \eqref{conventional C and P-2}, the action defined by ${\cal L}$ \eqref{Weinberg-type3} is transformed to
\begin{eqnarray}
\int d^{4}x \{\overline{\nu_{L}}(x)i\delslash \nu_{L}(x)
-(1/2)[-\nu_{L}^{T}(x)Cm^{\dagger}_{L}\nu_{L}(x) - \overline{\nu_{L}(x)}Cm_{L}\overline{\nu_{L}(x)}^{T}]\}
\end{eqnarray}
and thus ${\cal L}$ is CP invariant if  the symmetric mass matrix is pure imaginary  
\begin{eqnarray}\label{CP of mass term2}
m^{\dagger}_{L} = - m_{L}.
\end{eqnarray} 
To be precise, we use this CP invariance condition 
for the diagonalized mass eigenvalues, while leaving the bare parameter $m_{L}$ to be general and not necessarily pure imaginary. 
We diagonalize  the symmetric complex mass matrix by the $3\times 3$ Autonne-Takagi factorization using a unitary $U^{\prime}$ \cite{Autonne,Takagi}
\begin{eqnarray}\label{Autonne-Takagi2}
(U^{\prime})^{T}m_{L}U^{\prime} = iM   
\end{eqnarray}
with a real $3\times 3$ diagonal real matrix $M$ to ensure the Lagrangian after mass diagonalization is CP invariant in accord with the condition \eqref{CP of mass term2}.   We then define 
\begin{eqnarray}\label{A-T transformation2}
\nu_{L}(x)=U^{\prime}\tilde{\nu}_{L}(x).
\end{eqnarray}
Note that the unitary $U^{\prime}$ in \eqref{Autonne-Takagi2} is related to $U$ in \eqref{characteristic value1} by 
\begin{eqnarray}\label{U-prime}
U^{\prime} = U e^{i\pi/4}
\end{eqnarray} 
for a given $m_{L}$. Note that the Autonne-Takagi factorization is very different from the conventional diagonalization of a hermitian matrix by a unitary transformation; the Autonne-Takagi factorization basically gives rise to characteristic values but the phase freedom of the diagonal elements is still left free, and the specific phase convention in \eqref{Autonne-Takagi2} is chosen to define the CP invariant Lagrangian (using $\gamma^{0}$ parity) after the mass diagonalization.

After the mass diagonalization described above, we have a hermitian Lagrangian (suppressing the tilde-symbol of $\tilde{\nu}_{L}(x)$)
\begin{eqnarray}\label{Weinberg-type4}
{\cal L}
&=&\overline{\nu_{L}}(x)i\delslash \nu_{L}(x)
-(i/2)\{\nu_{L}^{T}(x)CM\nu_{L}(x)  - \overline{\nu_{L}(x)}CM\overline{\nu_{L}(x)}^{T}\}\nonumber\\
&=&(1/2)\{\overline{\psi}(x)i\delslash \psi(x)- \overline{\psi}(x)M\psi(x)\}          
\end{eqnarray}
where we defined
\begin{eqnarray}\label{Majorana in Weinberg2}
\psi(x) \equiv  e^{i\pi/4}\nu_{L}(x) +e^{-i\pi/4}C\overline{\nu_{L}}^{T}(x).
\end{eqnarray}
The hermitian Lagrangian \eqref{Weinberg-type4} is confirmed to be invariant under CP transformation in \eqref{conventional C and P-2} since real symmetric $M$ is chosen to satisfy \eqref{CP of mass term2}, and the effects of possible CP breaking contained in $U^{\prime}$ are transferred to the PMNS mixing matrix which contains a mixing matrix coming from the charged lepton sector also in an extension of the Standard Model.
The field $\psi(x)$ in \eqref{Majorana in Weinberg2} satisfies the classical Majorana condition
\begin{eqnarray}\label{classical condition2}
 C\overline{\psi(x)}^{T} = \psi(x)
\end{eqnarray}  
 identically regardless of the choice of $\nu_{L}(x)$.
 
The transformation \eqref{A-T transformation2} belongs to a canonical transformation which preserves the form of the kinetic term in the Lagrangian and thus preserves the canonical anti-commutation relations. In the canonical transformation, we apply the transformation rules of discrete symmetries  in \eqref{conventional C and P-2} to the new variables every time after the canonical transformation~\cite{Pauli, Gursey, KF-PG}.  The hermitian Lagrangian \eqref{Weinberg-type4} is thus  not invariant under C nor P in 
\eqref{conventional C and P-2} but invariant under the CP transformation since $M$ is real and symmetric, as we have already noted; 
  C and P  in \eqref{conventional C and P-2} are not assigned to the field $\nu_{L}$, but CP symmetry
$({\cal P}{\cal C})\nu_{L}(x)({\cal P}{\cal C})^{\dagger} = - \gamma^{0}C\overline{\nu_{L}(t,-\vec{x})}^{T}$ in \eqref{conventional C and P-2}
is well-defined. We thus naturally characterize the Majorana fermion \eqref{Majorana in Weinberg2} by the CP symmetry
\begin{eqnarray}
&&({\cal P}{\cal C}) (e^{i\pi/4}\nu_{L}(x) +e^{-i\pi/4}C\overline{\nu_{L}}^{T}(x))({\cal P}{\cal C})^{\dagger}\nonumber\\
&&= - i \gamma^{0}[e^{-i\pi/4}C\overline{\nu_{L}}^{T}(t,-\vec{x}) +e^{i\pi/4}\nu_{L}(t,-\vec{x})],
\end{eqnarray} 
namely,
\begin{eqnarray}\label{CP to define Majorana3}
 ({\cal P}{\cal C})\psi(x)({\cal P}{\cal C})^{\dagger}= -i\gamma^{0}C\overline{\psi(t,-\vec{x})}^{T}= - i\gamma^{0}\psi(t,-\vec{x}).
\end{eqnarray}
The first equality in \eqref{CP to define Majorana3} implies the operator relation while the second equality in \eqref{CP to define Majorana3}  implies the classical Majorana condition \eqref{classical condition2} which holds identically in the sense that \eqref{classical condition2} holds irrespective of the choice of $\nu_{L}(x)$. 

  Note that the operator relation in \eqref{CP to define Majorana3}
has a signature opposite to the relation $ ({\cal P}{\cal C})\psi(x)({\cal P}{\cal C})^{\dagger}= i\gamma^{0}\psi(t,-\vec{x})$ in \eqref{CP to define Majorana}, but it preserves the classical Majorana identity \eqref{classical condition2}
\begin{eqnarray}\label{CP of classical identity2}
 - i\gamma^{0}\psi(t,-\vec{x}) = C\overline{ - i\gamma^{0}\psi(t,-\vec{x})}^{T}
\end{eqnarray}
after the CP transformation. The CP symmetry and the classical Majorana condition are consistent.
When one defines the fermions $\psi(x)$ in terms of chiral fermions, which satisfy the classical Majorana condition $\psi(x)=C\overline{\psi(x)}^{T}$ identically, and if the action for the chiral fermions which determine Majorana fermions is invariant  under the CP symmetry as in \eqref{Weinberg-type2} and \eqref{Weinberg-type4}, the consistency condition is always satisfied: Namely, the CP transform of  Majorana fermions $\psi(x)$ satisfy the classical Majorana condition in the form either \eqref{CP of classical identity1} or \eqref{CP of classical identity2}.

The chiral component of the neutrino $\nu_{L}(x)$
 describes  the weak interaction in an extension of the Standard Model, which was originally defined in terms of the left-handed gauge eigenstates before the diagonalization of the neutrino masses,  
\begin{eqnarray}\label{Weak-coupling2}
&&\int d^{4}x[ (g/\sqrt{2})\bar{l}_{L}(x)\gamma^{\mu}W_{\mu}(x)U^{\prime}_{PMNS}\nu_{L}(x) + h.c.]\nonumber\\
&&= \int d^{4}x[ (g/\sqrt{2})\bar{l}_{L}(x)\gamma^{\mu}W_{\mu}(x)U_{PMNS}\hat{\nu}_{L}(x) + h.c.]
\end{eqnarray}
where we introduced an auxiliary  chiral variable  defined by 
\begin{eqnarray}
\hat{\nu}_{L}(x) \equiv e^{i\pi/4}\nu_{L}(x)
\end{eqnarray}
for which  the Majorana neutrino  \eqref{Majorana in Weinberg2}
is written as 
\begin{eqnarray}\label{Majorana with auxiliary}
\psi(x) = \hat{\nu}_{L}(x) + C\overline{\hat{\nu}_{L}}^{T}(x)
\end{eqnarray}
and many equations below are simplified.
The weak mixing matrix $U^{\prime}_{PMNS}$ is defined by $U^{\prime}$ in \eqref{U-prime} and thus $U^{\prime}_{PMNS}= U_{PMNS}e^{i\pi/4}$ which leads to $U^{\prime}_{PMNS}\nu_{L}(x)= U_{PMNS}\hat{\nu}_{L}(x)$ in \eqref{Weak-coupling2}. The mixing matrix $U_{PMNS}$ is defined in \eqref{Weak-coupling1}.
The C and P symmetries are not defined for $\hat{\nu}_{L}(x)$ without the right-handed components, but these symmetries are broken in the chiral weak interactions and thus need not be specified. 

The CP transformation law is 
\begin{eqnarray}\label{CP of hat-variable}
\hat{\nu}_{L}(x) = 
\frac{(1-\gamma_{5})}{2}\psi(x) \rightarrow 
\frac{(1-\gamma_{5})}{2}(-i\gamma^{0})\psi(t,-\vec{x}) =
-i\gamma^{0}C\overline{\hat{\nu}_{L}}^{T}(t, -\vec{x})
\end{eqnarray}
which contains an extra $U(1)$ phase $i=e^{i\pi/2}$ compared to the CP transformation law of $\nu_{L}(x)$ in \eqref{conventional C and P-2}. This CP transformation law is also directly obtained from $\nu_{L}(x) \rightarrow -\gamma^{0}C\overline{\nu_{L}(t,-\vec{x})}$ in \eqref{conventional C and P-2} and the definition $\nu_{L}(x) = e^{-i\pi/4}\hat{\nu}_{L}(x)$.  In accord with the transformation \eqref{CP of hat-variable}, it is convenient to formally assign the CP transformation law to the {\em charged lepton}
\begin{eqnarray}
l_{L}(x)  \rightarrow -i\gamma^{0}C\overline{l_{L}}^{T}(t, -\vec{x})
\end{eqnarray}
instead of the original $l_{L}(x)  \rightarrow -\gamma^{0}C\overline{l_{L}}^{T}(t, -\vec{x})$ in the analysis of CP breaking using the pair of fields $\{ l_{L}(x),\ \hat{\nu}_{L}(x)\}$. This formal replacement of CP law for the charged lepton does not change the CP analysis by \eqref{Weak-coupling2},   since the fermion number preserving charged-lepton sector after the mass diagonalization is invariant even when the extra overall $U(1)$ phase $i$ is added to the transformation law. Also, \eqref{Weak-coupling2} is invariant under the CP laws thus defined if one sets $U_{PMNS}=1$, and thus the modified CP laws are consistent to analyze the CP breaking induced by $U_{PMNS}$.

In this setting, the analysis of CP symmetry breaking is the same as the CP analysis with $i\gamma^{0}$-parity \eqref{Weak-coupling1} except for the replacement $i\gamma^{0} \rightarrow 
-i\gamma^{0}$ in the CP transformation laws, which does not change physics.  The absence of the fermion number symmetry in the neutrino sector \eqref{Weinberg-type4} accounts for the possible increase of the CP violating phases for the Majorana neutrinos \cite{Takasugi}.

Alternatively, one may consider the replacement in 
\eqref{Weak-coupling2}
\begin{eqnarray}\label{Weak-coupling}
&&\int d^{4}x[ (g/\sqrt{2})\bar{l}_{L}(x)\gamma^{\mu}W_{\mu}(x)U^{\prime}_{PMNS}\nu_{L}(x) + h.c.]\nonumber\\
&&\rightarrow \int d^{4}x[ (g/\sqrt{2})\bar{l}_{L}(x) \gamma^{\mu}W_{\mu}(x)U_{PMNS}\nu_{L}(x) + h.c.].
\end{eqnarray}
The chiral fermion $\nu_{L}(x)$ together with  $U^{\prime}_{PMNS} = U_{PMNS}e^{i\pi/4}$ describe weak interactions. But the overall $U(1)$
phase $e^{i\pi/4}$ in $U^{\prime}_{PMNS}$ is absorbed in the re-definition of the charged lepton fields, $\bar{l}_{L}(x)e^{i\pi/4} \rightarrow \bar{l}_{L}(x)$, and we obtain the last expression of \eqref{Weak-coupling} in the analysis of CP breaking. In this setting, we use the transformation rules of C,  P ($\gamma^{0}$-parity) and CP  for the chiral components of charged leptons and neutrinos in a uniform manner, in particular, the CP transformation
\begin{eqnarray}\label{common CP}
l_{L}(x) \rightarrow -\gamma^{0}C\overline{l_{L}}^{T}(t,-\vec{x}), \ \ \ \nu_{L}(x) \rightarrow -\gamma^{0}\overline{\nu_{L}}^{T}(t,-\vec{x}),
\end{eqnarray}
although some of them are broken.  The absence of the fermion number symmetry in the neutrino sector \eqref{Weinberg-type4} accounts the possible increase of the CP violating phases with the Majorana neutrinos \cite{Takasugi}. This formulation \eqref{Weak-coupling} appears to be a natural one in the actual analysis of the CP symmetry with the $\gamma^{0}$-parity. Physically, this formulation is equivalent to the case of the $i\gamma^{0}$-parity convention in \eqref{Weak-coupling1} in the analysis of CP breaking induced by $U_{PMNS}$.

The entire weak interaction is thus described by the chiral component  $\nu_{L}(x)$ using its CP property defined by the $\gamma^{0}$-parity.
 The Majorana neutrinos characterized by the CP symmetry retain certain information of their original chiral contents, for example, the parity by itself is not specified in the present characterization of  Majorana neutrinos (P-parity is not specified but CP-parity  is specified instead). 
 The fact that the neutrino is a chiral projection of the Majorana fermion  is assured by $\hat{\nu}_{L}(x)= [(1-\gamma_{5})/2]\psi(x) = e^{i\pi/4}\nu_{L}(x)$ and  \eqref{CP to define Majorana3} which contains the classical Majorana identity \eqref{classical condition2}.

If one wishes to define the parity and charge conjugation operators valid for the Majorana neutrino \eqref{Majorana in Weinberg2} in the present formulation,   it is possible to define a deformed symmetry generated by  \cite{KF-PG, Fujikawa}
\begin{eqnarray} \label{Deformed C and P-2} 
{\cal C}_{M}=1, \ \ \ {\cal P}_{M}={\cal P}{\cal C},
\end{eqnarray}
which is a symmetry of \eqref{Weinberg-type4} and
\begin{eqnarray}\label{trivial C and CP2}
{\cal C}_{M}\psi(x){\cal C}_{M}^{\dagger}=\psi(x),\ \ \ {\cal P}_{M}\psi(x){\cal P}_{M}^{\dagger}= -i\gamma^{0}\psi(t,-\vec{x}).
\end{eqnarray}
The non-trivial part of 
 this deformation is the CP symmetry and in this sense, this deformation is essentially equivalent to the formulation of the Majorana neutrino with ${\cal P}{\cal C}={\cal P}_{M}{\cal C}_{M}$ described above. The classical Majorana condition $\psi(x)=C\overline{\psi(x)}^{T}$, that determines if a given fermion is a Majorana fermion or not, carries the same physical information as the trivial operation ${\cal C}_{M}\psi(x){\cal C}_{M}^{\dagger}=\psi(x)$ applied to the fermion $\psi(x)$ which is assumed to be the Majorana fermion $\psi(x)=C\overline{\psi(x)}^{T}$.

The formulation \eqref{trivial C and CP2} with a deformed symmetry  leads to a formal enhancement of discrete symmetries in  \eqref{Weinberg-type4} by assigning C and P to the chiral component $\hat{\nu}_{L}(x)=(\frac{1-\gamma_{5}}{2})\psi(x) = e^{i\pi/4}\nu_{L}(x)$ in \eqref{Majorana with auxiliary},
\begin{eqnarray} \label{Majorana basis3} 
&&{\cal C}_{M}\hat{\nu}_{L}(x){\cal C}_{M}^{\dagger}=\hat{\nu}_{L}(x) =C\overline{\hat{\nu}_{R}}^{T}(x), \nonumber\\ 
&&{\cal P}_{M}\hat{\nu}_{L}(x){\cal P}_{M}^{\dagger}=-i\gamma^{0}\hat{\nu}_{R}(t,-\vec{x})
\end{eqnarray}
where we defined $\hat{\nu}_{R}(x)
\equiv (\frac{1+\gamma_{5}}{2})\psi(x) = C\overline{\hat{\nu}_{L}}^{T}(x)$, and 
\begin{eqnarray}\label{Majorana basis4}
{\cal C}_{M}\hat{\nu}_{R}(x){\cal C}_{M}^{\dagger}=C\overline{\hat{\nu}_{L}}^{T}(x), \ \  {\cal P}_{M}\hat{\nu}_{R}(x){\cal P}_{M}^{\dagger}=-i\gamma^{0}\hat{\nu}_{L}(t,-\vec{x}).
\end{eqnarray}
These transformation rules are mathematically consistent and imply perfect left-right symmetry  expected for a Majorana fermion $\psi=\hat{\nu}_{L}+\hat{\nu}_{R}$.

These rules \eqref{Majorana basis3} and \eqref{Majorana basis4} may be compared to the rules in \eqref{conventional C and P}. The physical degrees of freedom of a Majorana fermion $\psi=\hat{\nu}_{L}+\hat{\nu}_{R}$ are the same as either a left-handed chiral fermion or a right-handed chiral fermion but not both, contrary to the case of a Dirac fermion in \eqref{conventional C and P} where the left- and right-handed components are independent.  If one measures the left-handed projection of the Majorana fermion $\psi$, for example, one obtains the chiral freedom $\hat{\nu}_{L}$ and simultaneously the information of $\hat{\nu}_{R}$ also. Physically, this introduction of $ {\cal C}_{M}$ and $ {\cal P}_{M}$ does not add new ingredients to the analysis of weak interactions \cite{Fujikawa}.

\section{Discussion and conclusion}

We have demonstrated the physical equivalence of $\gamma^{0}$-parity and $i\gamma^{0}$-parity in the description of emergent Majorana neutrinos formed from chiral fermions using Weinberg's model of neutrinos in an extension of the Standard Model. In this analysis, the characterization of the emergent Majorana neutrinos by the CP symmetry plays an important role. We can describe the model for two different definitions of parity by the same physical parameters $M$ and $U_{PMNS}$, which are fixed by experiments,  using the chiral fermion $\nu_{L}(x)$ or $\hat{\nu}_{L}(x)$, which is defined as a chiral component such as  $\hat{\nu}_{L}(x)= [(1-\gamma_{5})/2]\psi(x)$ of  the Majorana field $\psi(x)$ in \eqref{Majorana with auxiliary}.  The analysis of CP breaking using the two different definitions of parity is thus equivalent. 
 It is known that Weinberg's model of Majorana neutrinos represent the essential aspects of the general models of Majorana neutrinos such as the seesaw models \cite{Xing, Fukugita, Giunti, Bilenky}, where two classes of massive Majorana neutrinos generally appear and both of them may be characterized by the CP symmetry. Our analysis thus justifies the use of either $\gamma^{0}$-parity or $i\gamma^{0}$-parity in the analysis of the Majorana neutrinos in an extension of SM. In Appendix, an explicit analysis of seesaw models is presented.

Theoretically, the two different parity operations are equivalent in the Standard Model where the  fermion number is conserved.  The Standard Model deformed by the lepton number violating neutrino mass term is still characterized by  the discrete symmetries C, P and CP in 
\eqref{conventional C and P} or \eqref{conventional C and P-2} after the diagonalization of the mass term by a canonical transformation. Neither C nor P is a good symmetry, but the combined CP symmetry can be always chosen to be a good symmetry after the mass diagonalization using the specific property of the Autonne-Takagi factorization.  
 The CP transformation of the emergent Majorana fermions thus defined is  consistent with the classical Majorana condition, as in \eqref{CP of classical identity1} and \eqref{CP of classical identity2}.
 The definition of Majorana fermions formed from chiral fermions by the CP symmetry works for either form of parity operation, by resolving the strictures implied by the original Majorana conditions on the parity operation in \eqref{reality condition} and \eqref{modified parity}.

It is also possible to reformulate the Majorana fermions defined by the CP symmetry using a formally deformed symmetry formed by ${\cal C}_{M}$ and ${\cal P}_{M}$ in \eqref{Deformed C and P} or in \eqref{Deformed C and P-2}, and it illustrates some of the naively expected  properties of the Majorana fermion and its chiral projections, as in \eqref{trivial C and CP} and \eqref{trivial C and CP2}. It is assuring that the ``parity'' represented by ${\cal P}_{M}$ is always the $\pm i\gamma^{0}$-parity to be consistent with the classical Majorana condition, irrespective of the definitions of parity for the starting fermions.  It has been noted, however, that this formulation with the deformed symmetry does not add new physical ingredients beyond the original definition of Majorana neutrinos by the CP symmetry \cite{Fujikawa}.    

 In connection with the deformed symmetry generated by ${\cal C}_{M}$ and ${\cal P}_{M}$ in \eqref{trivial C and CP}, one may consider another  approach which may  presumably be more common in neutrino physics including the seesaw models \cite{Xing, Fukugita, Giunti, Bilenky}. One may start with a simple diagonalization of  the given Lagrangian  \eqref{Weinberg-type} without asking C and P properties of the chiral fermion $\nu_{L}(x)$ by obtaining
 the Lagrangian \eqref{Weinberg-type2},
\begin{eqnarray}\label{Weinberg-type5}
{\cal L}
&=&\overline{\nu_{L}}(x)i\delslash \nu_{L}(x)
-(1/2)\{\nu_{L}^{T}(x)CM\nu_{L}(x) + h.c.\}\nonumber\\
&=&(1/2)\{\overline{\psi}(x)i\delslash \psi(x)- \overline{\psi}(x)M\psi(x)\}          
\end{eqnarray}
with
\begin{eqnarray} \label{Majorana from chiral-3}
\psi(x) = \nu_{L}(x)+ C\overline{\nu_{L}}^{T}(x)
\end{eqnarray}
  which satisfies the classical Majorana condition $\psi(x)=C\overline{\psi(x)}^{T}$ identically.  It is known that the essence of various seesaw models is covered by the above model \eqref{Weinberg-type5}.
After the considerations of several physical consistency conditions in \cite{Bilenky}, for example,  one arrives at the CP transformation law 
\begin{eqnarray}\label{CP for seesaw}
(\tilde{{\cal P}} \tilde{{\cal C}})\psi(x)(\tilde{{\cal P}} \tilde{{\cal C}})^{\dagger}=i\gamma^{0}\psi(t,-\vec{x})
\end{eqnarray} 
which agrees with our identification \eqref{CP to define Majorana}. 
In the common treatment of Majorana neutrinos in seesaw models in an extension of the Standard Model, it is customary to introduce a new charge conjugation law for Majorana neutrinos in \eqref{Majorana from chiral-3} \cite{Xing, Fukugita, Giunti, Bilenky}
\begin{eqnarray}
\tilde{{\cal C}}\psi(x)\tilde{{\cal C}}^{\dagger}=\psi(x)
\end{eqnarray}
with 
\begin{eqnarray}
\tilde{{\cal C}}\nu_{L}(x)\tilde{{\cal C}}^{\dagger}= C\overline{\nu_{L}(x)}^{T}
\end{eqnarray}
which has been named as the {\em pseudo C-symmetry} in \cite{FT1}. This rule may be compared to the standard C symmetry ${\cal C}\nu_{L}(x){\cal C}^{\dagger} = C\overline{\nu_{R}(x)}^{T}$ in \eqref{conventional C and P}. It is also possible to define the {\em pseudo P-symmetry} by \cite{Fujikawa} 
\begin{eqnarray}\label{parity1}
\tilde{{\cal P}}\psi(x)\tilde{{\cal P}}^{\dagger}=i\gamma^{0}\psi(t,-\vec{x})
\end{eqnarray}
with 
\begin{eqnarray}
\tilde{{\cal P}}\nu_{L}(x)\tilde{{\cal P}}^{\dagger}=i\gamma^{0}\nu_{L}(t,-\vec{x})
\end{eqnarray}
which may be compared to the standard parity
${\cal P}\nu_{L}(x){\cal P}^{\dagger}=i\gamma^{0}\nu_{R}(t,-\vec{x})$ in \eqref{conventional C and P}.
One can confirm that the CP operator defined by 
$\tilde{{\cal P}}\tilde{{\cal C}}$ generates the desired CP transformation in \eqref{CP for seesaw}  and $\nu_{L}(x) \rightarrow i\gamma^{0}C\overline{\nu_{L}(t,-\vec{x})}^{T}$. It may appear that $\tilde{{\cal C}}$ and $\tilde{{\cal P}}$ thus defined provide alternatives to ${\cal C}_{M}$ and ${\cal P}_{M}$ in \eqref{trivial C and CP}.

A drawback to the pseudo C and P operators defined above is that they are not operatorially consistent by noting 
$\nu_{L}(x)=(\frac{1-\gamma_{5}}{2})\nu_{L}(x)$ \cite{FT1,FT2, Fujikawa}
\begin{eqnarray}\label{pseudo operators}
&&\tilde{{\cal C}}\nu_{L}(x)\tilde{{\cal C}}^{\dagger}=(\frac{1-\gamma_{5}}{2})\tilde{{\cal C}}\nu_{L}(x)\tilde{{\cal C}}^{\dagger}=(\frac{1-\gamma_{5}}{2}) C\overline{\nu_{L}(x)}^{T}=0,\nonumber\\
&&\tilde{{\cal P}}\nu_{L}(x)\tilde{{\cal P}}^{\dagger}=(\frac{1-\gamma_{5}}{2})\tilde{{\cal P}}\nu_{L}(x)\tilde{{\cal P}}^{\dagger}=(\frac{1-\gamma_{5}}{2})i\gamma^{0}\nu_{L}(t,-\vec{x})=0
\end{eqnarray}
since $C\overline{\nu_{L}(x)}^{T}$ and $i\gamma^{0}\nu_{L}(t,-\vec{x})$ are both right-handed. The pseudo C and P symmetries cannot be substitutes for ${\cal C}_{M}$ and ${\cal P}_{M}$ in \eqref{trivial C and CP}. It has been emphasized in \cite{Fujikawa} that essentially the entire physics of Majorana neutrinos in an extension of SM is described by the CP symmetry without referring to C and P separately and thus the above operatorial inconsistency of pseudo C and P symmetry operators  has not been recognized in the past practical analyses of Majorana neutrinos \cite{Xing, Fukugita, Giunti, Bilenky}, as long as the basic CP transformation law \eqref{CP for seesaw} is correctly chosen.

Finally, we make a brief comment on the parity operation for the ``elementary'' Majorana fermion
in \eqref{reality condition} and \eqref{modified parity}, which motivated us to define the $i\gamma^{0}$-parity but has not played a direct role in the present analysis.  It is instructive to recall the construction of two Majorana fermions with a degenerate mass from a Dirac fermion to understand the role of parity in Majorana fermions. One may start with 
\begin{eqnarray}\label{Dirac action0}
{\cal L} &=& \int d^{4}x \overline{\psi_{D}(x)}[i\gamma^{\mu}\partial_{\mu} -m]\psi_{D}(x)\nonumber\\
&=& \int d^{4}x \overline{\psi_{+}(x)}[i\gamma^{\mu}\partial_{\mu} -m]\psi_{+}(x) + \int d^{4}x \overline{\psi_{-}(x)}[i\gamma^{\mu}\partial_{\mu} -m]\psi_{-}(x)
\end{eqnarray}
with 
\begin{eqnarray}
\psi_{+}=\frac{1}{\sqrt{2}}[\psi_{D}(x) + C\overline{\psi_{D}(x)}^{T}], \ \ \ \psi_{-}=\frac{1}{\sqrt{2}}[\psi_{D}(x) - C\overline{\psi_{D}(x)}^{T}],
\end{eqnarray}
which satisfy the classical Majorana condition
\begin{eqnarray}
C\overline{\psi_{+}(x)}^{T} = \psi_{+}(x), \ \ \ 
C\overline{\psi_{-}(x)}^{T} = -\psi_{-}(x) 
\end{eqnarray} 
and  the charge conjugation
\begin{eqnarray}
{\cal C} \psi_{+}(x) {\cal C}^{\dagger} = \psi_{+}(x), \ \ \ {\cal C} \psi_{-}(x) {\cal C}^{\dagger} = - \psi_{-}(x)
\end{eqnarray}
with $ {\cal C} \psi_{D}(x) {\cal C}^{\dagger}=C\overline{\psi_{D}(x)}^{T} = \psi_{D}^{c}(x)$ and $ {\cal C} \psi_{D}^{c}(x) {\cal C}^{\dagger}=\psi_{D}(x)$.  The classical Majorana condition and the  charge conjugation condition agree in this special case. Under the charge conjugation, the action \eqref{Dirac action0} is invariant.

If one adopts the $i\gamma^{0}$-parity ${\cal P}\psi_{D}(x){\cal P}^{\dagger} = i\gamma^{0}\psi_{D}(t,-\vec{x})$, one finds
\begin{eqnarray}\label{P and CP5-1}
&&{\cal P}\psi_{+}(x){\cal P}^{\dagger} = i\gamma^{0}\psi_{+}(t,-\vec{x}), \ \ \ {\cal P}\psi_{-}(x){\cal P}^{\dagger} = i\gamma^{0}\psi_{-}(t,-\vec{x}), \nonumber\\
&&{\cal P}{\cal C}\psi_{+}(x){\cal C}^{\dagger}{\cal P}^{\dagger} = i\gamma^{0}\psi_{+}(t,-\vec{x}), \ \ \ {\cal P}{\cal C}\psi_{-}(x){\cal C}^{\dagger}{\cal P}^{\dagger} = -i\gamma^{0}\psi_{-}(t,-\vec{x}),
\end{eqnarray}
while if one adopts the $\gamma^{0}$-parity ${\cal P}\psi_{D}(x){\cal P}^{\dagger} = \gamma^{0}\psi_{D}(t,-\vec{x})$, one finds
\begin{eqnarray}\label{P and CP5-2}
&&{\cal P}\psi_{+}(x){\cal P}^{\dagger} = \gamma^{0}\psi_{-}(t,-\vec{x}), \ \ \ {\cal P}\psi_{-}(x){\cal P}^{\dagger} = \gamma^{0}\psi_{+}(t,-\vec{x}), \nonumber\\
&&{\cal P}{\cal C}\psi_{+}(x){\cal C}^{\dagger}{\cal P}^{\dagger} = -\gamma^{0}\psi_{-}(t,-\vec{x}), \ \ \ {\cal P}{\cal C}\psi_{-}(x){\cal C}^{\dagger}{\cal P}^{\dagger} = \gamma^{0}\psi_{+}(t,-\vec{x}).
\end{eqnarray}
The action \eqref{Dirac action0} is invariant under  both \eqref{P and CP5-1}
and \eqref{P and CP5-2}. But the difference is that the $i\gamma^{0}$-parity \eqref{P and CP5-1} maintains the two Lagrangians of Majorana fermions $\psi_{+}(x)$ and $\psi_{-}(x)$ {\em separately} invariant in the action \eqref{Dirac action0}, while the $\gamma^{0}$-parity \eqref{P and CP5-2} interchanges the two Lagrangians of Majorana fermions in the action \eqref{Dirac action0}. It is interesting that the $\gamma^{0}$-parity is represented as a doublet representation in both the decomposition of a Dirac fermion $\psi_{D}(x)$ into two Majorana fermions  as in \eqref{P and CP5-2} 
\begin{eqnarray}
\psi_{D}(x) = \psi_{+}(x) + \psi_{-}(x)
\end{eqnarray}
and the chiral decomposition as in \eqref{conventional C and P-2}
\begin{eqnarray}
\psi_{D}(x) =  \psi_{L}(x) + \psi_{R}(x).
\end{eqnarray}

If one would extract the first action for the Majorana fermion $\psi_{+}(x)$ or the second action for the Majorana fermion $\psi_{-}(x)$ in the second line of \eqref{Dirac action0} as a mathematical model of a Majorana fermion, one would have to adopt $i\gamma^{0}$-parity.  In such a case, one would need to give a physical meaning to the imaginary parity eigenvalues $\pm i$~\footnote{It is customary to assign real eigenvalues $\pm 1$ to the parity operation and formulate the parity selection rules based on them in the case of parity conserving interactions, such as in the charmonium phenomenology.  I thank J. Arafune for a helpful discussion on parity operation. See also \cite{Kayser1}.}.
\\

I thank A. Tureanu for helpful comments. The present work is supported in part by JSPS KAKENHI (Grant No.18K03633).
 \\

\appendix
\section{Comment on the spontaneous parity breaking}
 
The following issue was raised by an anonymous referee: The two different definitions of parity are obviously equivalent for Dirac fermions. One may then expect that the equivalence of these two parity operations is more transparent in the left-right symmetric models for generating  Majorana neutrinos, for example,  the model by 
Mohapatra and Senjanovic \cite{Mohapatra}. 

We here show that the model \cite{Mohapatra}, which is regarded as one of the possible versions of the seesaw model, has  common features and no essential differences from other seesaw models, whose physics aspects are described by Weinberg's model of Majorana neutrinos we use in the body of the paper. The analysis in this Appendix is thus regarded as an extension of the analysis in the body of the present paper to an explicit analysis of a class of seesaw models.

They start with a left-right symmetric model following Pati and Salam \cite{Mohapatra}. They then introduce a spontaneous parity breaking, which breaks the fermion number symmetry simultaneously.  In the starting theory without the fermion number breaking one can use either $\gamma^{0}$ parity or $i\gamma^{0}$ parity for all the fermions. When one adds the fermion number violating Higgs term, these two parity operations are not apparently equivalent. Nevertheless, one can make the fermion number violating term invariant under either  $\gamma^{0}$ parity or $i\gamma^{0}$ parity  by suitably assigning the parity transformation laws to the extra Higgs field which breaks the fermion number and parity spontaneously. 

In this setting and after the spontaneous symmetry breaking, they obtain the neutrino sector (we write only the relevant terms in our notation) \cite{Mohapatra}
\begin{eqnarray}\label{seesaw Lagrangian-A}
{\cal L}=\frac{1}{2}\overline{\psi_{L}}
i\gamma^{\mu}\partial_{\mu} \psi_{L} + \frac{1}{2}\overline{\psi_{R}}i\gamma^{\mu}\partial_{\mu} \psi_{R} -\frac{1}{2}\overline{\psi_{L}}\left(\begin{array}{cc}
            m_{N}& m_{e}\\
            m_{e}&0
            \end{array}\right)\psi_{R} + h.c.
\end{eqnarray}
where the large mass parameter $m_{N}$ arises from the spontaneous symmetry breaking and 
\begin{eqnarray}\label{seesaw variables}
\psi_{L}=\left(\begin{array}{c}
            C\overline{\nu_{R}}^{T}\\
            \nu_{L}
            \end{array}\right), \ \ \ \psi_{R}=\left(\begin{array}{c}
            \nu_{R}\\
            C\overline{\nu_{L}}^{T}
            \end{array}\right).
\end{eqnarray}
After the Autonne-Takagi factorization,
\begin{eqnarray}\label{Autonne-Takagi1-A}
U^{T}\left(\begin{array}{cc}
            m_{N}& m_{e}\\
            m_{e}&0
            \end{array}\right)U =\left(\begin{array}{cc}
            M_{1}& 0\\
            0&-M_{2}
            \end{array}\right)
            \end{eqnarray}
with a unitary $U$ and 
\begin{eqnarray}
\tilde{\psi_{R}}=U\psi_{R}, \ \ \ \tilde{\psi_{L}}=U^{\star}\psi_{L}
\end{eqnarray}
the above Lagrangian is written as 
\begin{eqnarray}\label{Majorana-Lagrangian-A}
{\cal L}&=&\frac{1}{2}\overline{\tilde{\psi_{L}}}
i\gamma^{\mu}\partial_{\mu} \tilde{\psi_{L}} + \frac{1}{2}\overline{\tilde{\psi_{R}}}i\gamma^{\mu}\partial_{\mu} \tilde{\psi_{R}} -\frac{1}{2}\overline{\tilde{\psi_{L}}}\left(\begin{array}{cc}
            M_{1}& 0\\
            0&-M_{2}
            \end{array}\right)\tilde{\psi_{R}} + h.c.\nonumber\\
 &=& \overline{\tilde{\nu_{R}}}i\gamma^{\mu}\partial_{\mu} \tilde{\nu_{R}}  -\frac{1}{2}\tilde{\nu_{R}}^{T}CM_{1}\tilde{\nu_{R}} + h.c
+ \overline{\tilde{\nu_{L}}}
i\gamma^{\mu}\partial_{\mu} \tilde{\nu_{L}}  +\frac{1}{2}\tilde{\nu_{L}}^{T}CM_{2}\tilde{\nu_{L}} +h.c \nonumber\\
&=&\frac{1}{2}\overline{\psi_{1}}[i\gamma^{\mu}\partial_{\mu}-M_{1}]\psi_{1} +\frac{1}{2}\overline{\psi_{2}}[i\gamma^{\mu}\partial_{\mu}-M_{2}]\psi_{2},
\end{eqnarray}
where we defined 
\begin{eqnarray}\label{Majorana field1-A}
\psi_{1}=\tilde{\nu_{R}} + C\overline{\tilde{\nu_{R}}}^{T}, \ \ \ \psi_{2}=\tilde{\nu_{L}} - C\overline{\tilde{\nu_{L}}}^{T}
\end{eqnarray}
	which satisfy the Majorana conditions (to be definite we assume $M_{1} \gg M_{2}$, namely, $M_{1}\sim m_{N}$ and $M_{2}\sim m_{e}^{2}/m_{N}$)
\begin{eqnarray}\label{Majorana condition-A}
&&	\psi_{1}(x)= C\overline{\psi_{1}}^{T}(x), \ \ \  \ [i\gamma^{\mu}\partial_{\mu}-M_{1}]\psi_{1}(x)=0 ,   \nonumber\\
&&	\psi_{2}(x)= - C\overline{\psi_{2}}^{T}(x), \ \ \ \ [i\gamma^{\mu}\partial_{\mu}-M_{2}]\psi_{2}(x)=0. 
\end{eqnarray}	
The present scheme is valid for a model with 3 generations using $6\times 6$ Autonne-Takagi matrix 
$U$ and $3\times 3$ diagonal mass matrices $M_{1}$ and $M_{2}$ \cite{Mohapatra, KF-PG}, although we use the notation of a single generation model in this Appendix.

We examine the C, P and CP properties of Majorana neutrinos thus defined (by omitting the tilde symbols of $\tilde{\nu}_{L}$ and $\tilde{\nu}_{R}$, for notational simplicity).\\
(i) $\gamma^{0}$ parity:
\begin{eqnarray}\label{gamma-0 parity}
&&{\cal C}\psi_{1}(x){\cal C}^{\dagger}= \nu_{L}(x) + C\overline{\nu_{L}}^{T}(x),\nonumber\\
&&{\cal P}\psi_{1}(x){\cal P}^{\dagger}=\gamma^{0}[\nu_{L}(t,-\vec{x}) - C\overline{\nu_{L}}^{T}(t,-\vec{x})],\nonumber\\
&&{\cal PC}\psi_{1}(x){\cal C}^{\dagger}{\cal P}^{\dagger}= \gamma^{0}[\nu_{R}(t,-\vec{x}) - C\overline{\nu_{R}}^{T}(t,-\vec{x})],\nonumber\\
\nonumber\\
&&{\cal C}\psi_{2}(x){\cal C}^{\dagger}= -\nu_{R}(x) + C\overline{\nu_{R}}^{T}(x),\nonumber\\
&&{\cal P}\psi_{2}(x){\cal P}^{\dagger}=\gamma^{0}[\nu_{R}(t,-\vec{x}) + C\overline{\nu_{R}}^{T}(t,-\vec{x})],\nonumber\\
&&{\cal PC}\psi_{2}(x){\cal C}^{\dagger}{\cal P}^{\dagger}= -\gamma^{0}[\nu_{L}(t,-\vec{x}) + C\overline{\nu_{L}}^{T}(t,-\vec{x})].
\end{eqnarray}
We observe that none of $C, P$ and $CP$ defined in terms of $\gamma^{0}$ parity is the good symmetry of Majorana neutrinos. We next examine\\
(ii)$i\gamma^{0}$ parity:
\begin{eqnarray}\label{i-gamma-0 parity}
&&{\cal C}\psi_{1}(x){\cal C}^{\dagger}= \nu_{L}(x) + C\overline{\nu_{L}}^{T}(x),\nonumber\\
&&{\cal P}\psi_{1}(x){\cal P}^{\dagger}= i\gamma^{0}[\nu_{L}(t,-\vec{x}) + C\overline{\nu_{L}}^{T}(t,-\vec{x})],\nonumber\\
&&{\cal PC}\psi_{1}(x){\cal C}^{\dagger}{\cal P}^{\dagger}= i\gamma^{0}[\nu_{R}(t,-\vec{x}) + C\overline{\nu_{R}}^{T}(t,-\vec{x})]~=i\gamma^{0}\psi_{1}(t,-\vec{x}),\nonumber\\
\nonumber\\
&&{\cal C}\psi_{2}(x){\cal C}^{\dagger}= -\nu_{R}(x) + C\overline{\nu_{R}}^{T}(x),\nonumber\\
&&{\cal P}\psi_{2}(x){\cal P}^{\dagger}= i\gamma^{0}[\nu_{R}(t,-\vec{x}) - C\overline{\nu_{R}}^{T}(t,-\vec{x})],\nonumber\\
&&{\cal PC}\psi_{2}(x){\cal C}^{\dagger}{\cal P}^{\dagger}= -i\gamma^{0}[\nu_{L}(t,-\vec{x}) - C\overline{\nu_{L}}^{T}(t,-\vec{x})]~=-i\gamma^{0}\psi_{2}(t,-\vec{x}),
\end{eqnarray}
namely, only the CP symmetry in terms of $i\gamma^{0}$ parity is a good symmetry, ${\cal PC}\psi_{1}(x){\cal C}^{\dagger}{\cal P}^{\dagger} =i\gamma^{0}\psi_{1}(t,-\vec{x})$ and ${\cal PC}\psi_{2}(x){\cal C}^{\dagger}{\cal P}^{\dagger} = -i\gamma^{0}\psi_{2}(t,-\vec{x})$.

What we learn from the model with spontaneous parity and fermion number breaking is summarized as follows:\\
1. We can define CP for the neutrinos only if we use 
$i\gamma^{0}$ parity, and 
C and P are all broken in the neutrino sector, just as the case of the Weyl neutrino in SM.  CP thus defined is consistent with the Majorana condition \eqref{Majorana condition-A}. \\
2. We can  thus analyze the CP breaking arising from PMNS matrix using $i\gamma^{0}$ parity (not  $\gamma^{0}$ parity) in the present model (or seesaw models). This analysis is essentially the same as the analysis of Weinberg's  models with $i\gamma^{0}$ parity in subsection 2.1 of the body of the present paper and also the analysis of seesaw models in \cite{Fujikawa}.\\

One may next ask if the crucial difficulty \eqref{gamma-0 parity} of the conventional $\gamma^{0}$ parity, which has been used by many authors in the past,  is resolved? This is the main subject of subsection 2.2  in the body of the present paper. 
In the present context of a seesaw model, the essential idea is to use the Autonne-Takagi factorization in the form (instead of \eqref{Autonne-Takagi1-A})
\begin{eqnarray}\label{Autonne-Takagi2-prime}
{U^{\prime}}^{T}\left(\begin{array}{cc}
            m_{N}& m_{e}\\
            m_{e}&0
            \end{array}\right)U^{\prime} =(-i)\left(\begin{array}{cc}
            M_{1}& 0\\
            0&M_{2}
            \end{array}\right)
            \end{eqnarray}
with a unitary $U^{\prime}$ and 
\begin{eqnarray}
\tilde{\psi_{R}}=U^{\prime}\psi_{R}, \ \ \ \tilde{\psi_{L}}={U^{\prime}}^{\star}\psi_{L}
\end{eqnarray}
the above Lagrangian \eqref{seesaw Lagrangian-A} is written as 
\begin{eqnarray}\label{Majorana-Lagrangian-A-2}
{\cal L}&=&\frac{1}{2}\overline{\tilde{\psi_{L}}}
i\gamma^{\mu}\partial_{\mu} \tilde{\psi_{L}} + \frac{1}{2}\overline{\tilde{\psi_{R}}}i\gamma^{\mu}\partial_{\mu} \tilde{\psi_{R}} +\frac{i}{2}\overline{\tilde{\psi_{L}}}\left(\begin{array}{cc}
            M_{1}& 0\\
            0&M_{2}
            \end{array}\right)\tilde{\psi_{R}} + h.c.\nonumber\\
 &=& \overline{\tilde{\nu_{R}}}i\gamma^{\mu}\partial_{\mu} \tilde{\nu_{R}}  +\frac{i}{2}\tilde{\nu_{R}}^{T}CM_{1}\tilde{\nu_{R}} -\frac{i}{2}\overline{\tilde{\nu_{R}}}CM_{1}
 \overline{\tilde{\nu_{R}}}^{T}\nonumber\\
&+& \overline{\tilde{\nu_{L}}}
i\gamma^{\mu}\partial_{\mu} \tilde{\nu_{L}}  -\frac{i}{2}\tilde{\nu_{L}}^{T}CM_{2}\tilde{\nu_{L}} +\frac{i}{2}\overline{\tilde{\nu_{L}}}CM_{2}
 \overline{\tilde{\nu_{L}}}^{T}\nonumber\\
&=&\frac{1}{2}\overline{\psi_{1}}[i\gamma^{\mu}\partial_{\mu}-M_{1}]\psi_{1} +\frac{1}{2}\overline{\psi_{2}}[i\gamma^{\mu}\partial_{\mu}-M_{2}]\psi_{2},
\end{eqnarray}
where we defined 
\begin{eqnarray}\label{Majorana field2-A}
\psi_{1}=e^{i\pi/4}\tilde{\nu_{R}} -e^{-i\pi/4} C\overline{\tilde{\nu_{R}}}^{T}, \ \ \ \psi_{2}=e^{i\pi/4}\tilde{\nu_{L}} + e^{-i\pi/4}C\overline{\tilde{\nu_{L}}}^{T}
\end{eqnarray}
	which satisfy the Majorana conditions (to be definite we assume $M_{1} \gg M_{2}$, namely, $M_{1}\sim m_{N}$ and $M_{2}\sim m_{e}^{2}/m_{N}$)
\begin{eqnarray}\label{Majorana condition2-A}
&&	\psi_{1}(x)=- C\overline{\psi_{1}}^{T}(x), \ \ \  \ [i\gamma^{\mu}\partial_{\mu}-M_{1}]\psi_{1}(x)=0 ,   \nonumber\\
&&	\psi_{2}(x)=  C\overline{\psi_{2}}^{T}(x), \ \ \ \ [i\gamma^{\mu}\partial_{\mu}-M_{2}]\psi_{2}(x)=0. 
\end{eqnarray}	
Note that the Autonne-Takagi factorization is very different from the conventional diagonalization of a hermitian matrix by a unitary transformation; the Autonne-Takagi factorization basically gives rise to characteristic values but the phase freedom of the diagonal elements is still left free, and the specific phase convention in \eqref{Autonne-Takagi2-prime} is chosen to define the CP invariant Lagrangian (using $\gamma^{0}$ parity) after the mass diagonalization. In this choice of factorization it is crucial that the classical Majorana condition $\psi_{2}(x)=  C\overline{\psi_{2}}^{T}(x)$ in \eqref{Majorana condition2-A}, for example,  is valid for any choice of the phase 
$\psi_{2}=e^{i\alpha}\tilde{\nu_{L}} + e^{-i\alpha}C\overline{\tilde{\nu_{L}}}^{T}$
in \eqref{Majorana field2-A}. 

We now examine the C, P and CP properties of Majorana neutrinos thus defined in \eqref{Majorana field2-A}  (by omitting the tilde symbols of $\tilde{\nu}_{L}$ and $\tilde{\nu}_{R}$, for notational simplicity).\\
(i) $\gamma^{0}$ parity:
\begin{eqnarray}\label{gamma-0 parity2}
&&{\cal C}\psi_{1}(x){\cal C}^{\dagger}= -e^{-i\pi/4}\nu_{L}(x) + e^{i\pi/4}C\overline{\nu_{L}}^{T}(x)=i\psi_{2}(x),\nonumber\\
&&{\cal P}\psi_{1}(x){\cal P}^{\dagger}=\gamma^{0}[e^{i\pi/4}\nu_{L}(t,-\vec{x}) +e^{-i\pi/4} C\overline{\nu_{L}}^{T}(t,-\vec{x})]=\gamma^{0}\psi_{2}(t,-\vec{x}),\nonumber\\
&&{\cal PC}\psi_{1}(x){\cal C}^{\dagger}{\cal P}^{\dagger}= i\gamma^{0}[e^{i\pi/4}\nu_{R}(t,-\vec{x}) - e^{-i\pi/4}C\overline{\nu_{R}}^{T}(t,-\vec{x})]=i\gamma^{0}\psi_{1}(t,-\vec{x}),\nonumber\\
\nonumber\\
&&{\cal C}\psi_{2}(x){\cal C}^{\dagger}= e^{-i\pi/4}\nu_{R}(x) + e^{i\pi/4}C\overline{\nu_{R}}^{T}(x)=-i\psi_{1}(x),\nonumber\\
&&{\cal P}\psi_{2}(x){\cal P}^{\dagger}=\gamma^{0}[e^{i\pi/4}\nu_{R}(t,-\vec{x}) - e^{-i\pi/4}C\overline{\nu_{R}}^{T}(t,-\vec{x})]=\gamma^{0}\psi_{1}(t,-\vec{x}),\\
&&{\cal PC}\psi_{2}(x){\cal C}^{\dagger}{\cal P}^{\dagger}= -i\gamma^{0}[e^{i\pi/4}\nu_{L}(t,-\vec{x}) + e^{-i\pi/4}C\overline{\nu_{L}}^{T}(t,-\vec{x})]= -i\gamma^{0}\psi_{2}(t,-\vec{x}).\nonumber
\end{eqnarray}
We observe that $CP$ defined in terms of $\gamma^{0}$ parity is the good symmetry of Majorana neutrinos, ${\cal PC}\psi_{1}(x){\cal C}^{\dagger}{\cal P}^{\dagger}= i\gamma^{0}\psi_{1}(t,-\vec{x})$ and ${\cal PC}\psi_{2}(x){\cal C}^{\dagger}{\cal P}^{\dagger}= -i\gamma^{0}\psi_{2}(t,-\vec{x})$, although $C$ and $P$ are separately broken for $M_{1}\neq M_{2}$. The CP thus defined is consistent with the classical Majorana condition \eqref{Majorana condition2-A} and thus we can formulate the weak interactions of Majorana neutrinos consistently using $\gamma^{0}$ parity.

We next examine\\
(ii)$i\gamma^{0}$ parity:
\begin{eqnarray}\label{i-gamma-0 parity2}
&&{\cal C}\psi_{1}(x){\cal C}^{\dagger}= -e^{-i\pi/4}\nu_{L}(x) + e^{i\pi/4}C\overline{\nu_{L}}^{T}(x)=i\psi_{2}(x),\nonumber\\
&&{\cal P}\psi_{1}(x){\cal P}^{\dagger}= i\gamma^{0}[e^{i\pi/4}\nu_{L}(t,-\vec{x}) - e^{-i\pi/4}C\overline{\nu_{L}}^{T}(t,-\vec{x})],\nonumber\\
&&{\cal PC}\psi_{1}(x){\cal C}^{\dagger}{\cal P}^{\dagger}= -\gamma^{0}[e^{i\pi/4}\nu_{R}(t,-\vec{x}) + e^{-i\pi/4}C\overline{\nu_{R}}^{T}(t,-\vec{x})],\nonumber\\
\nonumber\\
&&{\cal C}\psi_{2}(x){\cal C}^{\dagger}= e^{-i\pi/4}\nu_{R}(x) + e^{i\pi/4}C\overline{\nu_{R}}^{T}(x)=-i\psi_{1}(x),\nonumber\\
&&{\cal P}\psi_{2}(x){\cal P}^{\dagger}= i\gamma^{0}[e^{i\pi/4}\nu_{R}(t,-\vec{x}) + e^{-i\pi/4}C\overline{\nu_{R}}^{T}(t,-\vec{x})],\nonumber\\
&&{\cal PC}\psi_{2}(x){\cal C}^{\dagger}{\cal P}^{\dagger}= \gamma^{0}[e^{i\pi/4}\nu_{L}(t,-\vec{x}) -e^{-i\pi/4} C\overline{\nu_{L}}^{T}(t,-\vec{x})],
\end{eqnarray}
namely, none of $C, P$ and $CP$ in terms of $i\gamma^{0}$ parity is a good symmetry for $M_{1}\neq M_{2}$. The $i\gamma^{0}$ parity is not consistently formulated with the present choice of Autonne-Takagi factorization.

One may thus conclude that a general class of seesaw models for Majorana neutrinos is formulated using either $i\gamma^{0}$
parity or $\gamma^{0}$ parity with the consistent good CP symmetry if one chooses the suitable forms of the Autonne-Takagi factorization of the mass matrix, in agreement with the analysis of Weinberg's model in the body of the present paper.
\\

Incidentally, how to define the consistent nontrivial C for Majorana neutrinos  is another interesting issue in the seesaw models, as is mentioned in Section 3. The nontrivial charge conjugation C for the Majorana is  apparently broken in the above formulation of the seesaw model  for $M_{1}\neq M_{2}$ as in  \eqref{gamma-0 parity},  \eqref{i-gamma-0 parity}, \eqref{gamma-0 parity2} and \eqref{i-gamma-0 parity2}. The possible resolution of this issue has been discussed in detail in \cite{Fujikawa}, \cite{FT1} and \cite{FT2}.

\end{document}